\documentclass[12pt]{article}

\usepackage[body={17.5cm, 21cm},right=2cm]{geometry}

\usepackage{color}
\usepackage{graphicx}
\usepackage{epsf}

\usepackage{amsmath}
\usepackage{amsmath}
\usepackage{amssymb}
\usepackage{epsfig}
\usepackage{cite}
\usepackage{color,colordvi}
\newcommand{\be}{\begin{eqnarray}}
\newcommand{\ee}{\end{eqnarray}}
\newcommand{\bi}{\begin{itemize}}
\newcommand{\ei}{\end{itemize}}

\newcommand{\bx}{{\bf{x}}}

\newcounter{hran}

\def\MSbar{\relax\ifmmode\overline{\rm MS}\else{$\overline{\rm MS}${ }}\fi}

\def\d{\rm d}

 \def\vx{\vec{ x}} 
\def\vk{\vec{k}}
\def\vy{\vec{y}}

\def\tr{{\rm tr}}

\def\simlt{\stackrel{<}{{}_\sim}}

\numberwithin{equation}{section}
\begin{document}
\vspace{5mm}
\vspace{0.5cm}
\begin{center}

\def\thefootnote{\fnsymbol{footnote}}

{\large \bf 
The Four-point Correlator  in Multifield Inflation,\\
\vspace{0.5cm}	
the  Operator Product Expansion and the  Symmetries of de Sitter 
}
\\[2.5cm]
{\large  A. Kehagias$^{a}$ and A. Riotto$^{b}$}
\\[0.5cm]

\vspace{.3cm}
{\normalsize {\it  $^{a}$ Physics Division, National Technical University of Athens, \\15780 Zografou Campus, Athens, Greece}}\\

\vspace{.3cm}
{\normalsize { \it $^{b}$ Department of Theoretical Physics and Center for Astroparticle Physics (CAP)\\ 24 quai E. Ansermet, CH-1211 Geneva 4, Switzerland}}\\

\vspace{.3cm}


\end{center}

\vspace{2.5cm}

\hrule \vspace{0.3cm}
{\small  \noindent \textbf{Abstract} \\[0.3cm]
\noindent 
We study the multifield inflationary models where the 
cosmological perturbation is sourced by light scalar fields other than the inflaton. We exploit the operator product 
expansion and partly the symmetries present during the de Sitter epoch to characterize the
non-Gaussian four-point  correlator in the squeezed limit.  We point out  that the 
contribution to it from the intrinsic non-Gaussianity  of the light fields at horizon crossing can be larger 
than the usually studied contribution arising 
on superhorizon scales and it comes with a different shape. Our findings indicate that particular attention needs 
to be taken when studying the effects of the primordial NG on real observables, such as  the clustering of dark matter halos.

\vspace{0.5cm}  \hrule
\vskip 1cm

\def\thefootnote{\arabic{footnote}}
\setcounter{footnote}{0}



\baselineskip= 19pt

\newpage 
\tableofcontents

\newcommand{\fix}{\Phi(\mathbf{x})}
\newcommand{\fiLx}{\Phi_{\rm L}(\mathbf{x})}
\newcommand{\fiNLx}{\Phi_{\rm NL}(\mathbf{x})}
\newcommand{\fik}{\Phi(\mathbf{k})}
\newcommand{\fiLk}{\Phi_{\rm L}(\mathbf{k})}
\newcommand{\fiLkone}{\Phi_{\rm L}(\mathbf{k_1})}
\newcommand{\fiLktwo}{\Phi_{\rm L}(\mathbf{k_2})}
\newcommand{\fiLkthree}{\Phi_{\rm L}(\mathbf{k_3})}
\newcommand{\fiLkfour}{\Phi_{\rm L}(\mathbf{k_4})}
\newcommand{\fiNLk}{\Phi_{\rm NL}(\mathbf{k})}
\newcommand{\fiNLkone}{\Phi_{\rm NL}(\mathbf{k_1})}
\newcommand{\fiNLktwo}{\Phi_{\rm NL}(\mathbf{k_2})}
\newcommand{\fiNLkthree}{\Phi_{\rm NL}(\mathbf{k_3})}

\newcommand{\kernel}{f_{\rm NL} (\mathbf{k_1},\mathbf{k_2},\mathbf{k_3})}
\newcommand{\dirac}{\delta^{(3)}\,(\mathbf{k_1+k_2-k})}
\newcommand{\dirackonektwokthree}{\delta^{(3)}\,(\mathbf{k_1+k_2+k_3})}

\newcommand{\beq}{\begin{equation}}
\newcommand{\eeq}{\end{equation}}
\newcommand{\beqarr}{\begin{eqnarray}}
\newcommand{\eeqarr}{\end{eqnarray}}

\newcommand{\angk}{\hat{k}}
\newcommand{\angn}{\hat{n}}

\newcommand{\tfnow}{\Delta_\ell(k,\tau_0)}
\newcommand{\tf}{\Delta_\ell(k)}
\newcommand{\tfone}{\Delta_{\el\ell_1}(k_1)}
\newcommand{\tftwo}{\Delta_{\el\ell_2}(k_2)}
\newcommand{\tfthree}{\Delta_{\el\ell_3}(k_3)}
\newcommand{\tffour}{\Delta_{\el\ell_1^\prime}(k)}
\newcommand{\deltatilde}{\widetilde{\Delta}_{\el\ell_3}(k_3)}

\newcommand{\alm}{a_{\ell m}}
\newcommand{\almL}{a_{\ell m}^{\rm L}}
\newcommand{\almNL}{a_{\ell m}^{\rm NL}}
\newcommand{\almone}{a_{\el\ell_1 m_1}}
\newcommand{\almLone}{a_{\el\ell_1 m_1}^{\rm L}}
\newcommand{\almNLone}{a_{\el\ell_1 m_1}^{\rm NL}}
\newcommand{\almtwo}{a_{\el\ell_2 m_2}}
\newcommand{\almLtwo}{a_{\el\ell_2 m_2}^{\rm L}}
\newcommand{\almNLtwo}{a_{\el\ell_2 m_2}^{\rm NL}}
\newcommand{\almthree}{a_{\el\ell_3 m_3}}
\newcommand{\almLthree}{a_{\el\ell_3 m_3}^{\rm L}}
\newcommand{\almNLthree}{a_{\el\ell_3 m_3}^{\rm NL}}

\newcommand{\YLMstar}{Y_{L M}^*}
\newcommand{\Ylmstar}{Y_{\ell m}^*}
\newcommand{\Ylmstarone}{Y_{\el\ell_1 m_1}^*}
\newcommand{\Ylmstartwo}{Y_{\el\ell_2 m_2}^*}
\newcommand{\Ylmstarthree}{Y_{\el\ell_3 m_3}^*}
\newcommand{\Ylmstarfour}{Y_{\el\ell_1^\prime m_1^\prime}^*}
\newcommand{\Ylmstarfive}{Y_{\el\ell_2^\prime m_2^\prime}^*}
\newcommand{\Ylmstarsix}{Y_{\el\ell_3^\prime m_3^\prime}^*}

\newcommand{\YLM}{Y_{L M}}
\newcommand{\Ylm}{Y_{\ell m}}
\newcommand{\Ylmone}{Y_{\el\ell_1 m_1}}
\newcommand{\Ylmtwo}{Y_{\el\ell_2 m_2}}
\newcommand{\Ylmthree}{Y_{\el\ell_3 m_3}}
\newcommand{\Ylmfour}{Y_{\el\ell_1^\prime m_1^\prime}}
\newcommand{\Ylmfive}{Y_{\el\ell_2^\prime m_2^\prime}}
\newcommand{\Ylmsix}{Y_{\el\ell_3^\prime m_3^\prime}}

\newcommand{\comm}[1]{\textbf{\textcolor{rossos}{#1}}}
\newcommand{\lsim}{\,\raisebox{-.1ex}{$_{\textstyle <}\atop^{\textstyle\sim}$}\,}
\newcommand{\gsim}{\,\raisebox{-.3ex}{$_{\textstyle >}\atop^{\textstyle\sim}$}\,}

\newcommand{\jl}{j_\ell(k r)}
\newcommand{\jlfourone}{j_{\el\ell_1^\prime}(k_1 r)}
\newcommand{\jlfivetwo}{j_{\el\ell_2^\prime}(k_2 r)}
\newcommand{\jlsixthree}{j_{\el\ell_3^\prime}(k_3 r)}
\newcommand{\jlsix}{j_{\el\ell_3^\prime}(k r)}
\newcommand{\jlthree}{j_{\el\ell_3}(k_3 r)}
\newcommand{\jlthreetau}{j_{\el\ell_3}(k r)}

\newcommand{\Gaunt}{\mathcal{G}_{\el\ell_1^\prime \, \el\ell_2^\prime \, 
\el\ell_3^\prime}^{m_1^\prime m_2^\prime m_3^\prime}}
\newcommand{\Gaunttwo}{\mathcal{G}_{\el\ell_1^\prime \, \el\ell_2^\prime \, 
\el\ell_3}^{m_1^\prime m_2^\prime m_3}}
\newcommand{\Gauntstardef}{\mathcal{H}_{\el\ell_1 \, \el\ell_2 \, \el\ell_3}^{m_1 m_2 m_3}}
\newcommand{\Gauntstarone}{\mathcal{G}_{\el\ell_1 \, L \,\, \el\ell_1^\prime}
^{-m_1 M m_1^\prime}}
\newcommand{\Gauntstartwo}{\mathcal{G}_{\el\ell_2^\prime \, \el\ell_2 \, L}
^{-m_2^\prime m_2 M}}

\newcommand{\de}{{\rm d}}

\newcommand{\dangn}{d \angn}
\newcommand{\dangk}{d \angk}
\newcommand{\dangkone}{d \angk_1}
\newcommand{\dangktwo}{d \angk_2}
\newcommand{\dangkthree}{d \angk_3}
\newcommand{\dk}{d^3 k}
\newcommand{\dkone}{d^3 k_1}
\newcommand{\dktwo}{d^3 k_2}
\newcommand{\dkthree}{d^3 k_3}
\newcommand{\dkfour}{d^3 k_4}
\newcommand{\dallk}{\dkone \dktwo \dk}

\newcommand{\FT}{ \int  \! \frac{d^3k}{(2\pi)^3} 
e^{i\mathbf{k} \cdot \angn \tau_0}}
\newcommand{\planewave}{e^{i\mathbf{k \cdot x}}}
\newcommand{\dallkfourier}{\frac{\dkone}{(2\pi)^3}\frac{\dktwo}{(2\pi)^3}
\frac{\dkthree}{(2\pi)^3}}

\newcommand{\Bis}{B_{\el\ell_1 \el\ell_2 \el\ell_3}^{m_1 m_2 m_3}}
\newcommand{\Avbis}{B_{\el\ell_1 \el\ell_2 \el\ell_3}}

\newcommand{\los}{\mathcal{L}_{\el\ell_3 \el\ell_1 \el\ell_2}^{L \, 
\el\ell_1^\prime \el\ell_2^\prime}(r)}
\newcommand{\loszero}{\mathcal{L}_{\el\ell_3 \el\ell_1 \el\ell_2}^{0 \, 
\el\ell_1^\prime \el\ell_2^\prime}(r)}
\newcommand{\losone}{\mathcal{L}_{\el\ell_3 \el\ell_1 \el\ell_2}^{1 \, 
\el\ell_1^\prime \el\ell_2^\prime}(r)}
\newcommand{\lostwo}{\mathcal{L}_{\el\ell_3 \el\ell_1 \el\ell_2}^{2 \, 
\el\ell_1^\prime \el\ell_2^\prime}(r)}
\newcommand{\losfNL}{\mathcal{L}_{\el\ell_3 \el\ell_1 \el\ell_2}^{0 \, 
\el\ell_1 \el\ell_2}(r)}



\def\d{d}
\def\C{{\rm CDM}}
\def\me{m_e}
\def\te{T_e}
\def\ti{\tau_{\rm initial}}
\def\tci#1{n_e(#1) \sigma_T a(#1)}
\def\tr{\tau_r}
\def\dtr{\delta\tau_r}
\def\dd{\widetilde\Delta^{\rm Doppler}}
\def\dsw{\Delta^{\rm Sachs-Wolfe}}
\def\clsw{C_\ell^{\rm Sachs-Wolfe}}
\def\cldop{C_\ell^{\rm Doppler}}
\def\Dt{\widetilde{\Delta}}
\def\mut{\mu}
\def\vt{\widetilde v}
\def\hp{ {\bf \hat p}}
\def\sdv{S_{\delta v}}
\def\svv{S_{vv}}
\def\bvt{\widetilde{\bv}}
\def\delt{\widetilde{\delta_e}}
\def\cos{{\rm cos}}
\def\nn{\nonumber \\}
\def\bq{ {\bf q} }
\def\ba{ {\bf p} }
\def\bap{ {\bf p'} }
\def\bqp{ {\bf q'} }
\def\bp{ {\bf p} }
\def\bpp{ {\bf p'} }
\def\bk{ {\bf k} }
\def\bx{ {\bf x} }
\def\bv{ {\bf v} }
\def\qp{ p^{\mu}k_{\mu} }
\def\qpp { p^{\mu} k'_{\mu} }
\def\bgm{ {\bf \gamma} }
\def\bkp{ {\bf k'} }
\def\gq{ g(\bq)}
\def\gqp{ g(\bqp)}
\def\fp{ f(\bp)}
\def\h#1{ {\bf \hat #1}}
\def\fpp{ f(\bpp)}
\def\fz{f^{(\vec{0})}(p)}
\def\fpz{f^{(\vec{0})}(p')}
\def\f#1{f^{(#1)}(\bp)}
\def\fps#1{f^{(#1)}(\bpp)}
\def\dq{ {d^3\bq \over (2\pi)^32E(\bq)} }
\def\dqp{ {d^3\bqp \over (2\pi)^32E(\bqp)} }
\def\dpp{ {d^3\bpp \over (2\pi)^32E(\bpp)} }
\def\dtq{ {d^3\bq \over (2\pi)^3} }
\def\dtqp{ {d^3\bqp \over (2\pi)^3} }
\def\dtpp{ {d^3\bpp \over (2\pi)^3} }
\def\part#1;#2 {\partial#1 \over \partial#2}
\def\deriv#1;#2 {d#1 \over d#2}
\def\Done{\Delta^{(1)}}
\def\Dtwo{\widetilde\Delta^{(2)}}
\def\fone{f^{(1)}}
\def\ftwo{f^{(2)}}
\def\tg{T_\gamma}
\def\delpp{\delta(p-p')}
\def\delb{\delta_B}
\def\tc{\tau_0}
\def\DD{\langle|\Delta(k,\mu,\tau_0)|^2\rangle}
\def\DDL{\langle|\Delta(k=l/\tc,\mu)|^2\rangle}
\def\bkpp{{\bf k''}}
\def\kmkp{|\bk-\bkp|}
\def\kmkpsq{k^2+k'^2-2kk'x}
\def\tt{ \left({\tau' \over \tau_c}\right)}
\def\kt{ k\mu \tau_c}

%
%
%

\section{Introduction}
\noindent
Primordial non-Gaussianity (NG)  of the cosmological perturbations has become a crucial aspect of both observational predictions of inflationary early universe models and of 
present and future observational probes of the Cosmic Microwave Background (CMB) anisotropies and of the Large Scale Structure (LSS) \cite{reviewNG}. The main motivation is that 
detecting, or simply constraining, deviations from a Gaussian distribution of the primordial fluctuations generated during an inflationary epoch \cite{lrreview} allows to discriminate among different scenarios for the generation of the 
primordial perturbations.  Indeed, a non-vanishing primordial NG encodes a wealth of information allowing to break  the degeneracy between models that, at the level of the power spectra, 
might result to be indistinguishable. 
The degeneracy stems from the  fact that during a period of exponential acceleration with  Hubble rate $H$
all scalar fields
 with a mass smaller than $H$  are inevitably quantum-mechanically excited
with a final superhorizon  flat spectrum. The  comoving curvature perturbation, which provides the initial conditions
for the CMB anisotropies and for the LSS of the universe, may be 
 generated not only by the same scalar field driving inflation (the inflaton), but also  when  the isocurvature perturbation, which is  associated to the fluctuations of these light scalar fields,  is converted
into curvature perturbation after (or at the end) of inflation \cite{curvaton1,LW,curvaton3,rate,end,during}. One typical  example is provided by the so-called inhomogeneous decay rate scenario \cite{rate} where the field driving  inflation (the inflaton) decays  perturbatively with a decay rate 
$\Gamma$. The reheating temperature $T_r$ of the hot plasma produced by the inflaton decay products is of the order of  $(M_{\rm Pl} \Gamma)^{1/2}$. If the decay rate depends on some light fields which are fluctuating during inflation, then  the corresponding 
 large scale spatial variations of the decay rate
will induce a temperature anisotropy, $\delta T_r/T_r\sim \delta\Gamma/\Gamma$. 

Distinguishing different  shapes of the primordial three- (bispectrum) and four-point (trispectrum) correlators, {\it i.e.} their dependence of the  momentum wave-vectors in Fourier space, is of crucial importance. Different mechanisms to generate the
inflationary perturbations give rise to unique 
signals with specific  shapes, which thus probe different aspects of the physics of the primordial universe. For example, models in
which the curvature perturbation is generated by an initial isocurvature perturbation
develop  (some of the)  non-linearities on superhorizon scales. The  corresponding NG  is of the local type, that is  the NG part of the primordial curvature perturbation is a local function of the Gaussian part. In momentum space, the three-point function arising from such a  local NG is dominated by the so-called �squeezed� configuration, where one of the momenta is
much smaller than the other two ($k_1 \ll k_2 \sim k_3$). The squeezed limit of NG is also    particularly interesting from the
 observationally point of view because   it leads to pronounced effects  on the clustering of dark matter halos and to  strongly scale-dependent bias \cite{dalal}. 
 It is impressive that a future detection of a high level of primordial NG in the squeezed configuration will  rule out all standard single-field models of 
inflation, where the same field drives inflation and is responsible for the perturbations, since they all predict very tiny deviations from Gaussianity \cite{noi,Maldacena}.   

In Ref. \cite{us} we have taken the first step in trying to characterize the three- and the four-point inflationary correlators
when the curvature perturbation is generated by scalar fields other than the inflaton, the so-called multifield inflation (for the case in which there is only one degree of freedom, see Refs. \cite{others}). In particular,  we have studied the implications of the symmetries present during a de Sitter phase, that is scale invariance and special conformal symmetry. For instance, we have shown that, as a consequence of the conformal symmetries, the two-point cross-correlation of the light fields vanish if their conformal weights (essentially their masses in units of 
the Hubble rate)  are different. Furthermore, 
we have pointed out that the Operator Product Expansion (OPE) technique is very suitable to analyze two
 interesting limits: the squeezed limit of the three-point correlator and the collapsed limit of the four-point correlator. Despite the fact that the conformal symmetry does not fix the
 shape of the four-point correlators of the light NG fields, we have been able to compute it  
 in the collapsed limit. As we mentioned, both the resulting shapes are relevant from the observational point of view.

In this paper we take a step further and study which informations we can get on the squeezed limit of the four-point correlator. This is an interesting question as the four-point function  is not fixed by  by conformal 
invariance of the de Sitter stage. Once more, we will resort to the OPE technique in order to learn what we can say about such a configuration of the four-point correlator. 

One of the goals of these paper is to stress that the contribution to the trispectrum in the squeezed limit
coming from the NG of the light fields at horizon crossing have a different shape and the amplitude can be  larger than the trispectrum  generated on superhorizon scales (which is there 
even if the light fields are gaussian). This is somewhat contrary to the common belief spread in the literature whose large majority has focused
on the NG originated at  superhorizon scales.

The paper is organized as follows. In section 2  we will present a summary of what we know about the bispectrum 
and trispectrum of the light fields during inflation thanks to the symmetry properties of de Sitter and the OPE technique. This section contains
known material and the expert  reader  can jump directly to  sections 3 and 4 where we calculate the four-point correlator in the squeezed limit
using various arguments. Section 5 contains some quantitative estimates of the three- and four-point correlators from the light 
NG fields; finally section 5 contains our conclusions.

\section{Some general considerations about  non-Gaussianities}
We can characterize the cosmological perturbations through 
 the $\delta N$ formalism \cite{deltaN}, where the comoving curvature perturbation $\zeta$ 
on a uniform
energy density hypersurface at time $t_{\rm f}$ is, on sufficiently large scales, equal to the perturbation
in the time integral of the local expansion from an initial flat hypersurface ($t = t_{*}$)
to the final uniform energy density hypersurface. On sufficiently large scales, the local expansion
can be approximated quite well by the expansion of the unperturbed Friedmann
universe. Hence the curvature perturbation at time $t_{\rm f}$ can be expressed in terms of  the values of the relevant scalar
fields $\sigma^I(t_{*},\vec{x})$ at $t_{*}$

\be
\zeta(t_{\rm f},\vec{x})=N_I\sigma^I+\frac{1}{2!}N_{IJ}\sigma^I\sigma^J+\frac{1}{3!}N_{IKJ}
\sigma^I\sigma^J\sigma^K+\cdots, \label{zeta}
\label{deltan}
\ee
where $N_I$, $N_{IJ}$ and $N_{IJK}$ are the first,  second and third derivative, respectively, of the number of e-folds 
\be
N(t_{\rm f},t_{*},\vec{x})=\int_{t_{*}}^{t_{\rm f}}\,{\rm d}t\, H(t,\vec{x})
\ee
with respect to the 
field $\sigma^I$. From the expansion (\ref{deltan}) one can read off the $n$-point correlators. 
For instance, the three- and four-point correlators of the comoving curvature perturbation  are given by
\begin{eqnarray}
\langle \zeta_{\vk_1}\zeta_{\vk_2}
\zeta_{\vk_3}\rangle&=&B^{\rm n-un}_\zeta(\vk_1,\vk_2,\vk_3)+B^{\rm un}_\zeta(\vk_1,\vk_2,\vk_3),\nonumber\\
B^{\rm n-un}_\zeta(\vk_1,\vk_2,\vk_3)&=&N_I N_J N_K B^{IJK}_{\vec{k}_1\vec{k}_2\vec{k}_3},\nonumber\\
B^{\rm un}_\zeta(\vk_1,\vk_2,\vk_3)&=& N_I N_{JK}N_{L}\left(P^{IK}_{\vec{k}_1}P^{JL}_{\vec{k}_2}+2\,\,{\rm permutations} \label{zeta3}
\right)
\end{eqnarray}
and 
\begin{eqnarray}
\langle \zeta_{\vk_1}\zeta_{\vk_2}
\zeta_{\vk_3}\zeta_{\vk_4}\rangle&=&T^{\rm n-un}_\zeta(\vk_1,\vk_2,\vk_3,\vk_4)+T^{\rm un}_\zeta(\vk_1,\vk_2,\vk_3,\vk_4),\nonumber\\
T^{\rm n-un}_\zeta(\vk_1,\vk_2,\vk_3,\vk_4)&=&N_I N_J N_K N_L T^{IJKL}_{\vec{k}_1\vec{k}_2\vec{k}_3\vec{k}_4}
+N_{IJ} N_{K}N_{L}N_M\left(P^{IK}_{\vec{k}_1}B^{JLM}_{\vec{k}_{12}\vec{k}_3\vec{k}_4}+11\,\,{\rm permutations}
\right),\nonumber\\
T^{\rm un}_\zeta(\vk_1,\vk_2,\vk_3,\vk_4)&=&N_{IJ} N_{KL}N_{M}N_N\left(P^{IL}_{\vec{k}_{12}}P^{JM}_{\vec{k}_{1}}P^{KN}_{\vec{k}_{3}}
+11\,\,{\rm permutations}
\right)
\nonumber\\
&+&N_{IJK} N_{L}N_{M}N_N\left(P^{IL}_{\vec{k}_{1}}P^{JM}_{\vec{k}_{2}}P^{KN}_{\vec{k}_{3}}
+3\,\,{\rm permutations}
\right).
 \label{zeta4}
\end{eqnarray}
Here  we have defined $\vec{k}_{ij}=(\vec{k}_{i}+\vec{k}_{j})$ and 

\begin{eqnarray}
\langle\sigma_{\vec{k}_{1}}^I\sigma^J_{\vec{k}_{2}}\rangle&=&(2\pi)^3\delta({\vec{k}_{1}}+{\vec{k}_{2}})P^{IJ}_{\vec{k}_{1}}=
(2\pi)^3\delta({\vec{k}_{1}}+{\vec{k}_{2}})\delta^{IJ}\left(\frac{H^2}{2k_1^3}\right)_{k_1=aH},\nonumber\\
\langle\sigma_{\vec{k}_{1}}^I\sigma^J_{\vec{k}_{2}}\sigma^K_{\vec{k}_{3}}\rangle&=&(2\pi)^3\delta({\vec{k}_{1}}+{\vec{k}_{2}}+{\vec{k}_{3}})B^{IJK}_{\vec{k}_{1}\vec{k}_{2}\vec{k}_{3}},\nonumber\\
\langle\sigma_{\vec{k}_{1}}^I\sigma^J_{\vec{k}_{2}}\sigma^J_{\vec{k}_{3}}\sigma^L_{\vec{k}_{4}}\rangle&=&(2\pi)^3\delta({\vec{k}_{1}}+{\vec{k}_{2}}+{\vec{k}_{3}}+{\vec{k}_{4}})T^{IJKL}_{\vec{k}_{1}\vec{k}_{2}\vec{k}_{3}\vec{k}_{4}}.\label{kro} 
\end{eqnarray}
The three-point correlator (and similarly  the four-point one) of the comoving curvature perturbation is  the sum of two pieces

\begin{itemize}
\item 
One, proportional to the connected three-point correlator of the $\sigma^I$ fields,  is present when the   fields $\sigma^I$ are intrinsically NG at horizon crossing. We dub it (in  a loose way) the non-universal contribution. 

\item The second one, which we dub universal (some people come them gravitational),     is generated when the modes of 
the fluctuations are super-Hubble and is present even if the $\sigma^I$ fields are gaussian. Of course they depend on the mechanism
for the conversion of the isocurvature modes into the curvature modes.

\end{itemize}
It is fair to say that most of the attention in the literature has been devoted to the universal contributions.  For instance, in the large majority of the literature on NG, the  nonlinear parameters\footnote{The prime denotes correlators without the $(2\pi)^3\delta(\sum_i\vk_i)$ factors. 
}

\begin{eqnarray}
f_{\rm NL}&=&\frac{5}{12}\frac{\langle \zeta_{\vk_1}\zeta_{\vk_2}
\zeta_{\vk_3}\rangle^\prime}{P^\zeta_{\vk_1}P^\zeta_{\vk_2}},\,\,\,\,\,\,\,\,\,\,\,\,\,\,\,\,\,\,\,\,\,\,\,\,\,\,\,\,\,\,\,\,\,\,\,\,\,\,\,\,\,\,\,\,\,\,\,\,\,\,\,\,\,\,\,({\rm squeezed}:\,\,k_1\ll k_2\sim k_3),\nonumber\\
\tau_{\rm NL}&=&\frac{1}{4}\frac{\langle \zeta_{\vk_1}\zeta_{\vk_2}\zeta_{\vk_3}\zeta_{\vk_4}\rangle^\prime}{P^\zeta_{\vk_1}
P^\zeta_{\vk_3}P^\zeta_{\vk_{12}}},\,\,\,\,\,\,\,\,\,\,\,\,\,\,\,\,\,\,\,\,\,\,\,\,\,\,\,\,\,\,\,\,\,\,\,\,\,\,\,\,\,\,\,\,\,\,\,\,\,\,\,\,\,({\rm collapsed}:\,\,\vk_{12}\simeq  0), \label{tf2}\nonumber\\
2\tau_{\rm NL}+\frac{54}{25}g_{\rm NL}&=&\frac{\langle \zeta_{\vk_1}\zeta_{\vk_2}\zeta_{\vk_3}\zeta_{\vk_4}\rangle^\prime}{P^\zeta_{\vk_4}
\left(P^\zeta_{\vk_1}P^\zeta_{\vk_{2}}+2\,\,{\rm permutations }\right)},\,\,\,\,\,\,\,({\rm squeezed}:\,\,k_4\ll k_1,k_2,k_3), \label{tf3}
\end{eqnarray}
are expressed as

\begin{eqnarray}
f^{\rm un}_{\rm NL}&=&\frac{5}{6}\frac{N^IN_{IJ}N^J}{(N_IN^I)^2},\nonumber\\
\tau^{\rm un}_{\rm NL}&=&
\frac{N^IN_{JI}N^{JK}N_K}{(N_IN^I)^3},\nonumber\\
g^{\rm un}_{\rm NL}&=&\frac{25}{54}
\frac{N^IN^JN^K N_{IJK}}{(N_IN^I)^3},
\end{eqnarray}
with no reference to the contribution to the non-universal terms. This might be due to the fact that the non-universal contributions
to the connected correlators generated if the light fields are NG depend on the specific self-interactions of the light fields  and in principle little was known about their
magnitude and shapes. 

To show that neglecting the non-universal contributions might lead to non correct conclusions about NG, let us consider the following simple example  where the primordial density perturbations is produced just
after the end of inflation through the modulated decay scenario when the decay rate of the inflaton is a function of some light field $\sigma$ \cite{rate}, that is  $\Gamma=\Gamma(\sigma)$.
If we approximate the inflaton reheating by a sudden decay, we may find an analytic estimate of the density perturbation.
In the case of modulated reheating, the decay occurs on a spatial hypersurface with
variable local decay rate and hence local Hubble rate $H=\Gamma(\sigma)$. Before the 
inflaton decay, the oscillating inflaton field has a pressureless equation of state and there is no density
perturbation. The perturbed expansion reads
\be
\delta N_{\rm d}=-\frac{1}{3}\ln\left(\frac{\rho_{\rm d}}{\overline{\rho}_{\rm d}}\right).
\ee
Immediately after the decay  we have radiation and hence the curvature perturbation reads
\be
\zeta=\delta N_{\rm d}+\frac{1}{4}\ln\left(\frac{\rho_{\rm d}}{\overline{\rho}_{\rm d}}\right).
\ee
Eliminating $\delta N_{\rm d}$ and using the local Friedmann equation $\rho\sim H^2$,
 to determine the local density in terms of the local decay rate $\Gamma=\Gamma(\sigma)$, we have at the linear order
\be
\label{rate1}
\zeta=-\frac{1}{6}\,\ln\left(\frac{\delta\Gamma}{\Gamma}\right).
\ee
Taylor expanding this expression in powers of the fluctuation $\sigma$, one obtains

\begin{eqnarray}
f^{\rm un}_{\rm NL}&=&5\left(1-\frac{\Gamma''\Gamma}{\Gamma'^{2}}\right),\nonumber\\
g^{\rm un}_{\rm NL}&=&\frac{25}{54}\frac{N'''}{N'^{3}}=
\frac{50}{3}\left(2-3\frac{\Gamma''\Gamma}{\Gamma'^{2}}+\frac{\Gamma'''}{\Gamma'^{3}}\right).
\end{eqnarray}
Now, suppose that the function $\Gamma(\sigma)$ is of the exponential type, $\Gamma(\sigma(\vx,t))\sim e^{a \sigma(\vx,t)}$ 
with $a$ constant. This  is a rather natural possibility if, for instance, the light field is a string modulus  setting the amplitude of some coupling constant. If so, one immediately
concludes that all the universal contributions to the NG vanish, $f^{\rm un}_{\rm NL}=\tau^{\rm un}_{\rm NL}=g^{\rm un}_{\rm NL}=\cdots=0$.
This holds to any order in perturbation theory as the relation between the comoving curvature perturbation $\zeta$ and the light fluctuation $\sigma$ is linear:   $\zeta\sim  \ln\left(\delta\Gamma/\Gamma\right)\sim \sigma$ from Eq. (\ref{rate1}). 
Alternatively, suppose that the function $\Gamma(\sigma)$ gets its dependence on the light field from some Yukawa-type interaction with Yukawa coupling $Y=Y_0(1+\sigma/2M)$, with $M$ some high mass scale.
If so, $\Gamma(h)\simeq \Gamma_0(1+\sigma/2M)^2$ and we obtain negligible NG,  $f^{\rm un}_{\rm NL}= 5/2$ and 
$g^{\rm un}_{\rm NL}= 25/3$, being the current bounds $f_{\rm NL}\simlt 10^2$ \cite{kom} and $g_{\rm NL}\simlt 10^6$ \cite{gnl}. 

These  simple examples indicate that, at least a priori, one may not disregard the non-universal contributions unless there is a convincing  argument
  that they are subleading. In fact, it was  pointed out long ago that the non-universal contributions may be of the same order of magnitude as the universal ones \cite{bmr, zal,fr1,fr2}. 
  
From now on we will therefore devote  our attention to  non-universal contributions to NG in order to characterize them. Let us therefore go back to such contributions
and briefly summarize what we know about them.

\subsection{Non-Gaussianities and  the conformal symmetries of de Sitter}
Some progress has been made recently on the knowledge of the NG carried by the light fields during the de Sitter stage\cite{anto,Creminelli,us}. 
Even though the intrinsically NG contributions to the $n$-point correlators of the light fields are model-dependent, 
their forms in some specific configurations are dictated
by the conformal symmetry of the de Sitter stage. 
Let us recall some of the properties of the conformal symmetry in de Sitter. For a more complete description the reader is referred to Ref. \cite{us}.
Conformal invariance 
in three-dimensional space $\mathbb{R}^3$ is connected to the symmetry under the group $SO(1,4)$ in the same way 
conformal invariance in a four-dimensional Minkowski spacetime is connected to the $SO(2,4)$ group. As $SO(1,4)$ is the 
isometry group of de Sitter spacetime,  a conformal phase during which fluctuations were generated could be 
a  de Sitter stage.  In such a  case, the kinematics  is specified by the embedding 
of $\mathbb{R}^3$ as flat sections in de Sitter spacetime. The de Sitter isometry group acts  as conformal group 
on $\mathbb{R}^3$ when the fluctuations are super-Hubble. It is in this regime that  the $SO(1,4)$ isometry
of the de Sitter background is realized as conformal symmetry of the flat $\mathbb{R}^3$ sections.  Correlators are expected to be constrained by conformal invariance. All these reasonings apply in the case in which the cosmological perturbations are generated 
by light scalar fields other than the inflaton. Indeed, it is only in such a case that correlators
inherit all the isometries of de Sitter. 
First, let us remind how  the conformal group  acts on super-Hubble scales. The set of transformations is given by 
\be
&& x_i'=a_i+M_i^{\,\,j}x_j, \\
&& x_i'=\lambda x_i, \\
&&x_i'=\frac{x_i+b_ix^2}{1+2b_ix_i+b^2x^2} \label{specconf}
\ee 
on Euclidean $\mathbb{R}^3$ with coordinates $x^i$.  These transformations correspond to translations and rotations (generated by $P_i,L_{ij}$), dilations
(generated by $D$) and special conformal transformations (generated by $K_i$), respectively, acting now on the constant time hypersurfaces  
of de Sitter spacetime.   It should be noted that special conformal transformations can be written in terms of inversion
\be
\label{inv}
x_i\to x_i'=\frac{x_i}{x^2} \label{inv}
\ee
as  (inversion)$\times$(translation)$\times$(inversion).
Under conformal transformations, the two-point function of fields $\sigma^I$ and $\sigma^J$ of 
conformal dimensions $\Delta_I$ and $\Delta_J$ respectively, transforms as 
\be
\langle\sigma^I(\vec{{x}}_1)\sigma^J(\vx_2)\rangle\to \Big{|}\frac{\partial x_i'}{\partial x_j}\Big{|}_{x=x_1}^{\Delta_I/3} 
\Big{|}\frac{\partial x'_i}{\partial x_j}\Big{|}_{x=x_2}^{\Delta_J/3} \langle\sigma^I(\vec{{x}}'_1)\sigma^J(\vx_2')\rangle
\ee
where $|\partial x_i' /\partial x_j|$ is the Jacobian of the transformation. For the  space inversion (\ref{inv}),
the two-point function, the form of which is forced by scale invariance,  transforms
as
\be
\langle\sigma^I(\vec{{x}}_1)\sigma^J(\vx_2)\rangle\to \frac{(x_1 x_2)^{\Delta_I+\Delta_J}}{x_1^{2\Delta_I}x_2^{2\Delta_J}}
\langle\sigma^I(\vec{{x}}_1)\sigma^J(\vx_2)\rangle,
\ee
 where for $\vx'=\vx/x^2$ we have used that   
\be
\left|\frac{\partial x_i'}{\partial x_j}\right|=\frac{1}{x^6}\, , ~~~x_{ij}\to \frac{x_{ij}}{x_i^2 x_j^2},
\label{spinv}
\ee
and the notation $x_{ij}=\left|\vx_i-\vx_j\right|$. Thus, space inversion leaves the two point function invariant if 
\be
\Delta_I=\Delta_J.
\ee
Similarly,  the three-point function transforms as 
\be
\langle\sigma^I(\vx_1)\sigma^J(\vx_2)\sigma^K(\vx_3)\rangle\to
\left|\frac{\partial x_i'}{\partial x_j}\right|_{x=x_1}^{\Delta_I/3} 
\left|\frac{\partial x_i'}{\partial x_j}\right|_{x=x_1}^{\Delta_J/3} 
\left|\frac{\partial x_i'}{\partial x_j}\right|_{x=x_1}^{\Delta_K/3} 
 \langle\sigma^I(\vx_1')\sigma^J(\vx_2')\sigma^K(\vx_3')\rangle
\ee
and using (\ref{spinv}), we get that the three-point correlator is invariant if 
\be
w_K=\Delta_I+\Delta_J-\Delta_K\, ,~~w_I=\Delta_J+\Delta_K-\Delta_I\, ,~~w_J=\Delta_I+\Delta_K-\Delta_J.
\ee 
As a result,  two- and three-point function are conformal invariant if they have the form
\be
\hspace{-1cm}&&\langle\sigma^I(\vec{{x}}_1)\sigma^J(\vx_2)\rangle=\left\{\begin{array}{cl}
                                                           \frac{c_{IJ}}{|\vx_1-\vx_2|^{2\Delta_I}}
&\Delta_I=\Delta_J,\label{2pc}\\
0 & \Delta_I\neq \Delta_J,
                                                          \end{array}
\right.
 \\
\hspace{-1cm}&&\langle\sigma^I(\vx_1)\sigma^J(\vx_2)\sigma^K(\vx_3)\rangle=
\frac{c_{IJK}}{|\vx_1-\vx_2|^{\Delta_I\!+\!\Delta_J\!-\!\Delta_K}|\vx_2-\vx_3|^{\Delta_J\!+\!\Delta_K\!-\!\Delta_I}
|\vx_3-\vx_1|^{\Delta_I\!+\!\Delta_K\!-\!\Delta_J}}, \label{3pc}
\ee
where again $\sigma^{I,J,K}$ are operators of dimensions $\Delta_{I,J,K}$. 
In other words, enhancing the symmetry including the special conformal symmetry has two consequences. First, the two-point functions are zero for 
operators with different dimensions and, second, the three-point functions are completely specified by special conformal 
transformations, {\it i.e.} by the full conformal symmetry. The four-point function on the other hand is not fixed by  by conformal 
invariance. However, as under special conformal symmetry we have

\be
x_{12}^{'2}=\frac{x_{12}^{2}}{\left|\vec{b}+\vx_1\right|^2\left|\vec{b}+\vx_2\right|^2},
\ee
the four-point correlation can be only a function of the cross-ratios $(x_{ij}x_{km}/x_{ik}x_{jm})$. The four-point correlator is therefore of the form

\be
\langle\sigma^I(\vx_1)\sigma^J(\vx_2)\sigma^K(\vx_3)\sigma^L(\vx_4)\rangle=F^{IJKL}\left(\frac{x_{12}x_{34}}{x_{13} x_{24}},
\frac{x_{14}x_{23}}{x_{13} x_{24}} \right)\prod_{i<j} x_{ij}^{\Delta/3-\Delta_I-\Delta_J} \label{44pt}
\ee
with $\Delta=\sum_I\Delta_I$. The four-point function is restricted but not fully specified by
conformal invariance to be a function of the so-called anharmonic ratios. 
Therefore, one can conclude that \cite{Creminelli,us}

\begin{itemize}

\item As a consequence of special
conformal symmetry 
 the scale invariance during the de Sitter stage, the two-point cross-correlation of the light fields vanish
 if their conformal weights  are different. Therefore, no assumption is needed on their  cross-correlation, 
it is simply dictated by the conformal symmetry \cite{us}. This is the reason why we inserted a Kronecker 
delta function $\delta^{IJ}$ in the expression (\ref{kro}).

\item The form of the three-point correlator $\langle\sigma_{\vec{k}_{1}}^I\sigma^J_{\vec{k}_{2}}\sigma^K_{\vec{k}_{3}}\rangle$
is fixed by conformal invariance of the de Sitter stage and in the squeezed limit it contributes to the total three-point correlator of the comoving curvature perturbation $\langle \zeta_{\vk_1}\zeta_{\vk_2}\zeta_{\vk_3}\rangle$ 
with the same shape of the universal contributions \cite{Creminelli,us}.

\item While the form of the four-point correlator $\langle\sigma_{\vec{k}_{1}}^I\sigma^J_{\vec{k}_{2}}\sigma^J_{\vec{k}_{3}}\sigma^L_{\vec{k}_{4}}\rangle$ is not fixed by the 
conformal symmetries of the de Sitter stage, in the collapsed limit it contributes to the total four-point correlator $\langle \zeta_{\vk_1}\zeta_{\vk_2}\zeta_{\vk_3}\zeta_{\vk_4}\rangle$ with the same shape of the universal contributions \cite{us}: it can be expressed as a product of power spectra.

\end{itemize}
In the rest of the paper we will characterize the non-universal contributions to the three- and four-point correlators expanding the results of Ref. \cite{us} and using  the OPE technique.

\subsection{Non-Gaussianities and the Operator product expansion}
\noindent
The OPE  is a  very  powerful tool to analyze the squeezed limit of the three-point correlator and the collapsed and squeezed limit of the four-point correlator. 
These limits 
 are  particularly interesting from the
 observational point of view because they are associated to the local model of NG  
 (for a review see  \cite{reviewNG}) which  leads to pronounced effects of NG on the clustering of dark matter halos and to  strongly scale-dependent bias \cite{dalal}. 
The OPE  has been established in perturbative quantum field theories. 
It is by now a standard tool in the analysis of 
gauge theories such as QCD and  Wilson's OPE \cite{Wilson} is the basis of virtually all calculations of nonperturbative effects in
analytical QCD.  
It is believed that all quantum field theories with well-behaved ultraviolet behavior  have
an OPE \cite{Wilson,wilson,zim}. 

Let us consider two generic operators $\sigma^I(\vx)$ and $\sigma^J(\vy)$ at the points $\vx$ and $\vy$ on a 
$\tau={\rm constant}$ hypersurface of de Sitter spacetime. 
Then, we expect that the product of  local operators are distances small compared to the 
characteristic length of the system should look like a local operator. As a result, we expect that
 the product of $\sigma^I(\vx)\sigma^J(\vy)$ of the two operators $\sigma^I(\vx)$ and $\sigma^J(\vy)$, located at nearby points $\vx$ and $\vy$, will have a 
short-distance expansion of the form  \cite{Wilson}
\be
\label{ab1}
\sigma^I(\vx) \sigma^J(\vy)\stackrel{\vx\to\vy}{\sim}\sum_nC_{n}(\vx-\vy;g){\cal{O}}_n(\vy),  
\ee
where $C_n(\vx-\vy)$ are c-number functions (in fact distributions), ${\cal{O}}_n$ local operators and $g$ is the coupling.                                                                                                  
Moreover, for $H\tau \ll 1$ we expect the OPE  to respect the symmetries of the de Sitter
spacetime realized non-linearly on the $\tau={\rm constant}$ hypersurface. Let us briefly summarize the results obtained in Ref. \cite{us}.

Let us note immediately that $n$-point correlators are reduced to calculation of three-point functions by repeated 
applications of  (\ref{ab1}). Let us consider the case where ${\cal{O}}$ is the field itself, {\it i.e.}
\be
\sigma^I(\vx) \sigma^J(\vy)\stackrel{\vx\to\vy}{\sim}\sum_K C^{IJ}_K(\vx-\vy;g)\sigma^K(\vy).  \label{ab2}
\ee
The $n$- and $(n+1)$-point functions are given by
\be
&&g_{n+1}^{I_1\cdots I_{n+1}}(x_1,\ldots,x_{n+1};\mu,g)=
\langle \sigma^{I_1}(x_1)\cdots\sigma^{I_{n+1}}(x_{n+1})\rangle',\\
&&g_{n}^{I_1\cdots I_{n}}(x_1,\ldots,x_{n};\mu,g)=
\langle \sigma^{I_1}(x_1)\cdots\sigma^{I_{n}}(x_{n})\rangle',
\ee
where $\mu$ a mass scale.
These correlators satisfy the Callan-Symanzik equation
\be
\left(\mu\frac{\partial}{\partial\mu}+\beta \frac{\partial}{\partial g}+\sum_I\gamma_I\right)g^{I}_i=0, ~~~~~~(i=n,n+1),
\ee
where $\beta$ is the usual $\beta$-function and $\gamma_I$ the anomalous dimension of $\sigma^I$. Using the OPE expansion (\ref{ab2})
one finds immediately that 
\be
g_{n+1}^{I_1\ldots I_{n+1}}=\sum_K C^{I_nI_{n+1}}_{K}g_{n}^{I_1\ldots I_{n-1}K}.
\ee
Then, the coefficients of the OPE expansion are also satisfy the  Callan-Symanzik equation
\be
\left(\mu\frac{\partial}{\partial\mu}+\beta \frac{\partial}{\partial g}+\gamma_I+\gamma_J-\gamma_K\right)C^{IJ}_K(x,y;\mu,g)=0.
\ee
In particular, for a a conformal field theory for which $\beta=0$, dimensional arguments and the fact that renormalized operators
can be chosen such that they do not depend  $\mu$ lead to   
\be
C^{IJ}_K(x,y;g)=\frac{C^{IJ}(g)}{x^{2 w_I+2w_J-2w_K}}, \label{c12}
\ee
where $w_{I,J,K}$ are the dimensions of the fields $\sigma^I$, $\sigma^J$ and $\sigma^K$, respectively. Therefore (\ref{ab2}) can be written as 
\be
\sigma^I(\vx) \sigma^J(\vy)\stackrel{\vx\to\vy}{\sim}\sum_K \frac{C^{IJ}(g)}{|\vx-\vy|^{2 w_I+2w_J-2w_K}}\sigma^K(\vy).
\ee
Here $C^{IJ}$ should be understood in a non-perturbative sense. In perturbation theory, it has an expansion 
in terms of the coupling(s)  $g$, {\it i.e.} for a single field $C(g)=c_0+c_1 g+c_2 g^2+\cdots$.  Let us now analyze in detail the three- and the four-point correlator in various interesting configurations.

\subsection{The three-point correlator in the squeezed limit}
\noindent
If we wish to consider the three-point correlator in the squeezed limit, the configuration in real space is such that two points, say
$\vx_1$ and $\vx_2$ are very close and the third one very far. 
Let us therefore consider the OPE expansion for the two fields $\sigma^I$  
and $\sigma^J$    in the (12) channel at the coincident point 
\be
\sigma^I(\vx_1)\sigma^J(\vx_2)\!=\!\left(\frac{C_0^{IJ}}{x_{12}^{2w}}+\frac{{C^{IJ}}_M}{x_{12}^w}\sigma^M(\vx_2)+\cdots\!\right).
\label{asd}
\ee
Here $w\simeq m^2/3 H^2\ll 1$,  where $m$ is the mass of the fields, is the conformal weight of the fields involved (remember that 
the weight of the fields $\sigma^I$ and $\sigma^I$  must be the same due to the special conformal symmetry). 
The three-point correlator in the squeezed limit  can be evaluated 
by employing the OPE  as
\be
\langle\sigma^I(\vx_1)\sigma^J(\vx_2)\sigma^K(\vx_3)\rangle=
\Big<\left(\frac{C^{IJ}_0}{x_{12}^{2w}}+\frac{C^{IJ}_A}{x_{12}^w}\sigma^A(\vx_2)+\cdots\right)
\sigma^K(\vx_3)\Big>.
\ee
Using again the orthogonality of the two-point functions,  one finds \cite{us}

\be
\langle\sigma^I(\vx_1)\sigma^J(\vx_2)\sigma^K(\vx_3)\rangle=\frac{{C^{IJ}}_A}{x_{12}^w}\langle\sigma^A(\vx_2)\sigma^K(\vx_3)\rangle
=\frac{C^{IJK}}{x_{12}^wx_{23}^{2w}}  ~~~(x_{12}\simeq 0). \label{3xv}
\ee
Using the expression
\be
\frac{1}{|\vx|^{w}}= 
\frac{\Gamma(\frac{3-w}{2})}{2^{w}\pi^{3/2}\Gamma(\frac{w}{2})}
\int {\rm d}^3 k\,|\vk|^{w-3}e^{-i\vk\cdot \vx},\label{xw}
\ee 
we obtain for an almost scale invariant  spectrum $w\approx 0$  the Fourier transform of Eq. (\ref{3xv})
\be
\langle\sigma^I_{\vec{k}_1}\sigma^J_{\vec{k}_2}\sigma^K_{\vec{k}_3}\rangle^\prime
\sim  C^{IJK}
   P_{\vk_1}P_{\vk_2}\left[1+{\cal O}\left(\frac{k_1^2}{k_2^2}\right)\right],\,\,\,\,\,\,(k_1\ll k_2\sim k_3).
\ee
The non-universal contribution to the three-point correlator in the squeezed limit has therefore the same shape of the universal contribution. Its amplitude 
is model-dependent.

\subsection{The four-point correlator in the collapsed limit}
\noindent
If we wish to consider the four-point correlator in the collapsed limit, the configuration in real space is such that two pairs of points, say
$\vx_1$, $\vx_2$ and $\vx_3$, $\vx_4$ are very far from each other. 
Let us therefore consider the OPE expansion (\ref{asd}) as well as the one for the other (34) channel at the coincident point
\be
\sigma^K(\vx_3)\sigma^L(\vx_4)\!=\!\left(\frac{C_0^{KL}(w)}{x_{34}^{2w}}+\frac{{C^{KL}}_M(w)}{x_{34}^w}\sigma^M(\vx_4)\!+\!\cdots\right).
 \label{34}
\ee
The four-point function in the collapsed limit 
\be
\langle\sigma^I(\vx_1)\sigma^J(\vx_2)\sigma^K(\vx_3)\sigma^L(\vx_4)\rangle ~~~~~~~~~(x_{12}\simeq 0\,\,{\rm and \,\,} x_{34}\simeq  0)
\ee 
can be expressed as

\be
\langle\sigma^I(\vx_1)\sigma^J(\vx_2)\sigma^K(\vx_3)\sigma^L(\vx_4)\rangle=\frac{C_0^{IJ}{C_0^{KL}}}{x_{12}^{2w}x_{34}^{2w}}+
\frac{{C^{IJ}}_A{C^{KL}}_B}{x_{12}^wx_{34}^w}\langle\sigma^A(\vx_2)\sigma^B(\vx_4)\rangle+\cdots, \label{ab}
\ee
whose Fourier transforms keeping the  connected contribution gives

 \be
 \label{asd}
 \langle \sigma_{\vk_1}^I\sigma_{\vk_2}^J\sigma_{\vk_3}^K\sigma_{\vk_4}^L\rangle^\prime \sim {C^{IJ}}_A C^{KLA}
P_{\vk_{12}}P_{\vk_2}P_{\vk_4}+\,\,{\rm permutations},\,\,\,\,\,\,\,\,\,\,(\vk_{12}\simeq \vec{0}).
\ee
The non-universal contribution to the four-point correlator in the collapsed limit has therefore the same shape of the universal contribution. Its amplitude 
is model-dependent.
\section{The four-point correlator in the squeezed limit}
\noindent
Let us now consider the four-point correlator in the squeezed limit
which was not analyzed in Ref. \cite{us}. We 
consider three points being  close to each other and the fourth very far apart.
In other words we consider the configuration  
\be
x_{14}\approx x_{24}\approx x_{34}\gg  x_{ij}, ~~~~(i,j=1,2,3). \label{xxx}
\ee
The  method  to characterize the four-point correlation in the squeezed limit is based entirely on the OPE.  As we said, we consider the  generic four-point correlator  in the squeezed limit in which one of the point is much
far from the remaining three, say $\ell = x_{12}\simeq x_{23} \simeq x_{13}\gg x_{14}$. We start by dividing space into volumes centered around the points $\vx_1,\vx_2$ and $\vx_3$. For simplicity we will take
all of these volumes to have the same shape and size $R\ll \ell $.  We wish to compute the correlator for the
smoothed functions 

\be
\sigma_R^I(\vx_i)=\int{\rm d}^3 x\, W_R(\vx-\vx_i)\, \sigma^I(\vx),\,\,\,\,\,\,(i=1,2,3),
\ee
where $W_R(\vx)$ is a window function selecting a volume of size $R$. The key point is that the leading term of the  four-point correlator in the squeezed limit can be computed considering the modulation of the three-point
correlator inside large volume with size at least $\ell$ where  
long wavelength fluctuation modes $\sigma^M_\ell(\vx)$ live (one can imagine to obtain them by smoothing with a 
window function $(1-W_R(\vx))$.  

The OPE for the two fields $\sigma_R^I$  
and $\sigma_R^J$    in the (12) channel at the coincident point is 
\be
&&\sigma_R^I(\vx_1)\sigma_R^J(\vx_2)\!=\!\frac{C_0^{IJ}(w)}{x_{12}^{2w}}+
\frac{{C^{IJ}}_M(w,g,{\sigma}^N_\ell(\vx_2))}{x_{12}^w}\sigma_R^M(\vx_2)+\cdots,
\label{12}
\ee
where we have made explicit that
 the coefficients of the expansion depend on coupling constants, denoted collectively by $g$  and on the long-wavelength mode 
${\sigma}^N_\ell(\vx_2)$ of the field ${\sigma}^N(\vx_2)$. Indeed, this fluctuation is seen by  the
remaining three short wavelength fluctuations as a vacuum  expectation value and it should be added to the zero mode
$\overline{\sigma}^N$.
Multiplying both sides with $\sigma_R^K(\vx_3)$ and since $x_{13}\approx 0$, we get
\be
&&\sigma_R^I(\vx_1)\sigma_R^J(\vx_2)\sigma_R^K(\vx_3)=\frac{C_0^{IJ}(w)}{x_{12}^{2w}}\sigma_R^K(\vx_3)+
\frac{{C^{IJ}}_M(w,g,{\sigma}^N_\ell(\vx_2))}{x_{12}^w}\sigma_R^M(\vx_2)\sigma_R^K(\vx_3)+\cdots.  \label{123}
\ee
We now expand the three-point correlator ${C^{IJ}}_M(w,g,{\sigma}^N_\ell(\vx_2))$ around $\overline{\sigma}^N$ in powers
of ${\sigma}^N_\ell(\vx_2)$ to  linear order

\be
\sigma_R^I(\vx_1)\sigma_R^J(\vx_2)\sigma_R^K(\vx_3)&=&\frac{C_0^{IJ}(w)}{x_{12}^{2w}}\sigma_R^K(\vx_3)+
\frac{{C^{IJ}}_M(w,g,\overline{\sigma}^N)}{x_{12}^w}\sigma_R^M(\vx_2)\sigma_R^K(\vx_3)\nonumber\\
&+&\frac{\partial{C^{IJ}}_M(w,g,\overline{\sigma}^N)}{\partial \overline{\sigma}^N}
\frac{\sigma_R^M(\vx_2)\sigma_R^K(\vx_3){\sigma}^N_\ell(\vx_2)}{x_{12}^w}
+\cdots  \label{123expanded}
\ee
and by applying the OPE in the (23) channel, we get

\be
\sigma_R^I(\vx_1)\sigma_R^J(\vx_2)\sigma_R^K(\vx_3)&=&\frac{C_0^{IJ}(w)}{x_{12}^{2w}}\sigma_R^K(\vx_3)+
\frac{{C^{IJ}}_M(w,g,\overline{\sigma}^N)}{x_{12}^w}\frac{C_0^{MK}(w)}{x_{23}^{2w}}\nonumber\\
&+&\frac{{C^{IJ}}_M(w,g,\overline{\sigma}^N)}{x_{12}^w}\frac{{C^{MK}}_P(w,g,\overline{\sigma}^N)}{x_{23}^w}\sigma_R^P(\vx_3)
\nonumber\\
&+&\frac{\partial{C^{IJ}}_M(w,g,\overline{\sigma}^N)}{\partial \overline{\sigma}^N}
\frac{\sigma_R^M(\vx_2)\sigma_R^K(\vx_3){\sigma}^N_\ell(\vx_2)}{x_{12}^w}
+\cdots.  \label{123expanded}
\ee
If we now multiply the above expression by $\sigma^L(\vx_4)$ and take the expectation value and Fourier transform, we finally obtain

\be
\fbox{$\displaystyle
\langle \sigma^{I}_{\vk_1}\sigma^{J}_{\vk_2}\sigma^K_{\vk_{3}}\sigma^L_{\vk_{4}}\rangle'\sim 
\frac{\partial{C^{IJ}}_M}{\partial \overline{\sigma}^L}\,C^{-1}_{NPM}
\,P_{\vk_4}\, \langle \sigma^{N}_{\vk_1}\sigma^{P}_{\vk_2}\sigma^K_{\vk_{3}}\rangle' 
+{C^{KL}}_M \,P_{\vk_4}\, \langle \sigma^{I}_{\vk_1}\sigma^{J}_{\vk_2}\sigma^M_{\vk_{3}}\rangle',\,\, (k_4\ll k_1,k_2,k_3)$}.
\label{4oplongt}
\ee
The non-universal four-point correlator (\ref{4oplongt}) can be therefore thought as the modulated three-point correlator in the presence of a long wavelength mode at the linear order and at the quadratic order in the coupling constants ${C^{IJ}}_M$ 

\be
\langle \sigma^{I}_{\vk_1}\sigma^{J}_{\vk_2}\sigma^K_{\vk_{3}}\sigma^L_{\vk_{4}}\rangle'\simeq \Big<\sigma^L_{\vk_4}
\langle \sigma^{I}_{\vk_1}\sigma^{J}_{\vk_2}\sigma^K_{\vk_{3}}\rangle'_{{\sigma}_\ell^L}
\Big>.
\label{mod}
\ee
One can therefore construct first the bispectrum and then replace the (fictitious, if necessary) zero mode vacuum expectation value, if any, with a long wavelength mode to be contracted to build up the four-point correlator. 
In particular, the piece linear in the coupling constant ${C^{IJ}}_M$ is similar to the $g^{\rm un}_{\rm NL}$ contribution, while the
quadratic piece in the coupling constant ${C^{IJ}}_M$ is similar to the $\tau^{\rm un}_{\rm NL}$ contribution. One should remember though that the ${C^{IJ}}_M$  is not only a tree-level quantity, but it is supposed to contain informations about  all loops.

The modulation
effect  in Eq. (\ref{4oplongt}) does not come as a surprise. Consider the contribution to the four-point correlator of the comoving curvature perturbation
coming from the universal terms in Eq. (\ref{zeta4}) in the squeezed limit $k_4\ll k_1,k_2,k_3$

\be
T^{\rm un}_\zeta(\vk_1,\vk_2,\vk_3,\vk_4)=\left(2N_{IJ} N_{KL}N_{M}N_N+N_{IJK} N_{L}N_{M}N_N\right)P^{IL}_{\vec{k}_{4}}\left(P^{JM}_{\vec{k}_{2}}P^{KN}_{\vec{k}_{3}}
+2 \,\,{\rm permutations}
\right)
 \label{zeta44}
\ee
and the contribution to the three-point correlator of the comoving curvature perturbation
coming from the universal terms in Eq. (\ref{zeta3})

\be
B^{\rm un}_\zeta(\vk_1,\vk_2,\vk_3)= N_I N_{JK}N_{L}\left(P^{IK}_{\vec{k}_1}P^{JL}_{\vec{k}_2}+2\,\,{\rm permutations} \label{zeta33}
\right).
\ee
Suppose now that there is long wavelength mode, associated to the fluctuation  $\sigma^M_{\ell}$ and to the wavenumber $\vk_4$, which modulates the bispectrum
(\ref{zeta33}). This   long wavelength mode is added to the zero mode $\overline{\sigma}^M$ from the point of view
of the short wavelength modes. One can Taylor expand the derivatives of the number of e-folds

\begin{eqnarray}
N_I(\overline{\sigma}^M+\sigma^M_{\ell})&\simeq& N_I(\overline{\sigma}^M)+N_{IM}(\overline{\sigma}^M)\sigma^M_{\ell},\nonumber\\
N_{IJ}(\overline{\sigma}^M+\sigma^M_{\ell})&\simeq& N_{IJ}(\overline{\sigma}^M)+N_{IJM}(\overline{\sigma}^M)\sigma^M_{\ell}
\label{10}
\end{eqnarray}
and immediately see that the modulation of the three-point correlator in the background of the  long wavelength mode gives
Eq. (\ref{zeta44}) in the squeezed limit

\be
T^{\rm un}_\zeta(\vk_1,\vk_2,\vk_3,\vk_4)\simeq \Big<\sigma^M_{\vk_4}
B^{\rm un}_\zeta\left(\sigma^M_{\vk_4};\vk_1,\vk_2,\vk_3\right)
\Big>.
\ee
Notice that by the same argument one can also reproduce the extra   contribution to the non-universal trispectrum
 in the squeezed limit
\be 
T^{\rm n-un}_\zeta(\vk_1,\vk_2,\vk_3,\vk_4)\supset N_{IJ} N_{K}N_{L}N_M\left(P^{IK}_{\vec{k}_1}B^{JLM}_{\vec{k}_{12}\vec{k}_3\vec{k}_4}+2\,\,{\rm permutations}
\right)
\ee
which comes from the long wavelength expansion (\ref{10}) of the term $
B^{\rm n-un}_\zeta(\vk_1,\vk_2,\vk_3)=N_I N_J N_K B^{IJK}_{\vec{k}_1\vec{k}_2\vec{k}_3}$.

We would like to stress out at this point that the non-universal contribution to the four-point correlator (\ref{4oplongt}) is not of the same form of the universal one  (\ref{zeta44}). Indeed, in the expression  (\ref{4oplongt}) the bispectrum is not evaluated in the squeezed limit and therefore it is not generically written as the products of power spectra. In fact, this is also true for the four-point correlator in the collapsed limit: the expression  (\ref{asd}) is value only in the limit $\vk_{12}\simeq \vec{0}$, but is second order in the coupling constants. There might well be another piece which does not diverge when $\vk_{12}\simeq \vec{0}$, but that is first order in the coupling constant. Which one dominates
clearly depends on the magnitude of such coupling. We will return to this point in the next subsection.

\subsection{A consistency check}
To check the validity of the expression (\ref{mod}), let us imagine to have only one test field $\sigma(\vx)$ with potential
$V(\sigma)=\lambda\sigma^p/p!$, with $p$ some positive integer larger than three. Let us also suppose that the 
light field has a vacuum expectation value $\overline{\sigma}$ which induces an interaction of the form $\lambda\overline{\sigma}\sigma^{p-1}/(p-1)!$ (for $p=4$ there will be therefore both a trispectrum and a bispectrum).
The $n$-th correlator point 
is given by 

\be
\langle\sigma_{\vk_1}(\tau)\sigma_{\vk_2}(\tau)\cdots \sigma_{\vk_n}(\tau)\rangle
=-i\Big< 0\left|\int_{-\infty}^\tau\,{\rm d}\tau'\,\left[\sigma_{\vk_1}(\tau)\sigma_{\vk_2}(\tau)\cdots \sigma_{\vk_n}(\tau),V(\tau')
\right] 0\right|\Big>.
\ee
Using the mode functions in de Sitter

\be
\sigma_{\vk}(\tau)=\frac{H}{\sqrt{2k^3}}\left(1-ik\tau\right)e^{-ik\tau},
\ee
one obtains \cite{zal}

\be
\langle\sigma_{\vk_1}\sigma_{\vk_2}\cdots \sigma_{\vk_n}\rangle'=V^{(n)}\frac{H^{2n-4} 
\left(k^{(n)}_t\right)^3}{\Pi_i 2 k_i^3}\, I_n(k_1,k_2,\cdots,k_n),  \label{n-pt}
\ee
where 
$k^{(n)}_t=k_1+k_2+\cdots k_n$, $V^{(n)}={\rm d}^n  V(\overline{\sigma})/{\rm d}\overline{\sigma}^n$ and 
\be
I_n(k_1,k_2,\cdots,k_n)=2\int_{-\infty}^{\tau_{\rm end}} \frac{{\rm d}\tau}{k_t^3\tau^4}{\rm{Re}}\Big{[}-i(1-ik_1 \tau)
\cdots(1-ik_n\tau)e^{ik^{(n)}_t\tau}\Big{]}.
\ee
In the squeezed limit 
$k_n\ll k_1,k_2,\cdots, k_{n-1}$ we have  $k_n\tau\ll k_i\tau$ for $(i=1,\ldots,n-1)$. Hence,  we find that 
\be
I_n=I_{n-1} +{\cal{O}}\left(\frac{k_n}{k^{(n)}_t}\right)
\ee
and the expression (\ref{n-pt}) turns out to be
\be
\langle\sigma_{\vk_1}\sigma_{\vk_2}\cdots \sigma_{\vk_n}\rangle'&=&V^{(n)}\frac{H^{2n-4} 
\left(k^{(n)}_t\right)^3}{\Pi_i 2 k_i^3}\, I_n(k_1,k_2,\cdots,k_n)\nonumber \\
&\simeq& V^{(n)}\frac{H^{2n-4} 
\left(k^{(n-1)}_t\right)^3}{2 k_n^3\Pi_{i}^{n-1} 2 k_i^3}\, I_{n-1}(k_1,k_2,\cdots,k_{n-1})\nonumber \\
&=&\frac{V^{(n)}}{V^{(n-1)}}\frac{H^2}{2k_n^3}\langle\sigma_{\vk_1}\sigma_{\vk_2}\cdots \sigma_{\vk_{n-1}}\rangle'\nonumber\\
&=&\frac{V^{(n)}}{V^{(n-1)}}P_{\vk_{n}}\langle\sigma_{\vk_1}\sigma_{\vk_2}\cdots \sigma_{\vk_{n-1}}\rangle'
\ee
In particular, for a potential $V(\sigma)=\lambda \sigma^4/4!$, we find that   
\begin{eqnarray}
I_3(k_1,k_2,k_3)&=&\frac{8}{9}-\frac{\sum_{i<j} 2 k_i k_j}{\left(k^{(n)}_t\right)^2}-\frac{1}{3}
\left(\gamma_{\rm E}+N_{k_t}\right)\frac{\sum_{i} 2 k^3_i}{\left(k^{(3)}_t\right)^3},\,\nonumber\\
I_4(k_1,k_2,k_3,k_4)&=&\frac{8}{9}-\frac{\sum_{i<j} 2 k_i k_j}{\left(k^{(n)}_t\right)^2}+2\frac{\Pi_{i}  k_i }{k_t^4}
-\frac{1}{3}\left(\gamma_{\rm E}+N_{k_t}\right)\frac{\sum_{i} 2 k^3_i}{\left(k^{(4)}_t\right)^3}.
\end{eqnarray}
Here $\gamma_{\rm E}$ is the Euler gamma and $N_{k_t}$ is the number of e-folds from the time the mode $k_t$ crosses the Hubble radius
to the time of end of inflation at $\tau_{\rm end}$. As a result, the four-point correlator is
\begin{eqnarray}
\langle\sigma_{\vk_1}\sigma_{\vk_2}\sigma_{\vk_3} \sigma_{\vk_4}\rangle'&=&\lambda\, \frac{H^4 \left(k^{(4)}_t\right)^3}{16\,k_1^3\,k_2^3\,k_3^3\,k_4^3}I_4(k_1,k_2,k_3,k_4), 
\end{eqnarray}
which, in the squeezed limit $k_4\ll k_1,k_2,k_3$   reduces to 

\begin{eqnarray}
\langle\sigma_{\vk_1}\sigma_{\vk_2}\sigma_{\vk_3} \sigma_{\vk_4}\rangle'&=&\lambda\, \frac{H^4 \left(k^{(4)}_t\right)^3}{16\,k_1^3\,k_2^3\,k_3^3\,k_4^3}I_4(k_1,k_2,k_3,k_4)\nonumber\\
&\simeq &\lambda\, \frac{H^4 \left(k^{(3)}_t\right)^3}{16\,k_1^3\,k_2^3\,k_3^3\,k_4^3}I_3(k_1,k_2,k_3)\nonumber\\
&=&\frac{1}{{\overline{\sigma}}}\, \frac{H^2}{2\,k_4^3}\langle\sigma_{\vk_1}\sigma_{\vk_2}\sigma_{\vk_3}\rangle'\nonumber\\
&=& \frac{P_{\vk_4}}{\overline{\sigma}}\langle\sigma_{\vk_1}\sigma_{\vk_2}\sigma_{\vk_3}\rangle', \label{zz}
\end{eqnarray}
and reproduces the expression the first piece of the expression (\ref{mod}) since in the example at hand we have  $\ln\, {C^{IJ\cdots K}}=\ln \, \overline{\sigma}+$ constant.
It is not difficult to see that the same conclusion holds at second order in the coupling constant ${C^{IJ\cdots K}}$ (or $\lambda$) by constructing
the bispectrum at this order through the interaction $\lambda\overline{\sigma}\sigma^3$ and  $\lambda\overline{\sigma}^2\sigma^2$, see Fig. 1.
\begin{figure}[h!]
\begin{center}
 \includegraphics[scale=0.4]{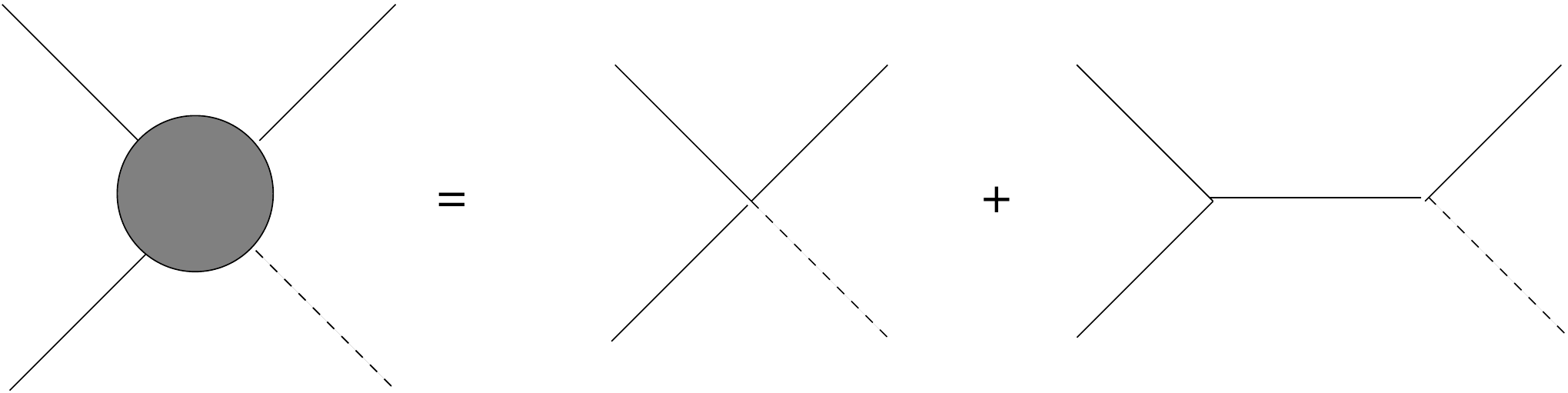}
\caption{Tree diagrams contributing to the modulation of the three-point function of short wavelength modes (continues lines)
in the background of a long wavelength mode (dashed lines).}
\end{center} 
\end{figure} 
From the example described above one can also read the information that the four-point correlator in the squeezed limit
is not generically written as the products of power spectra, as the universal contribution is. Furthermore, at the linear order
in the coupling constant $\lambda$ the four-point correlator in the collapsed limit does not have the form acquired
by the universal piece. This happens only at the order ${\cal O }(\lambda^2)$. Which form is the dominate one depends therefore on the
value of the coupling $\lambda$.

\section{The four-point correlator in the squeezed limit:  alternative methods}
In this section we wish to offer two alternative methods to get the  non-universal four-point correlator in the squeezed limit.
They are based both on the the symmetries of de Sitter and on the OPE technique.

\subsection{First method}
The fact that the four-point correlator in the squeezed limit  may be obtained from the three-point correlator can be
 also suggested by inspection of the general form (\ref{44pt}) dictated by the symmetries of de Sitter. Taking the limit
of  
Eq. (\ref{xxx}), it is easy to see that Eq. (\ref{44pt}) takes the form
\be
\langle\sigma^I(\vx_1)\sigma^J(\vx_2)\sigma^K(\vx_3)\sigma^L(\vx_4)\rangle=\frac{1}{x_{14}^{2w}}\frac{1}{(x_{12}x_{23}x_{13})^{2w/3}}
F^{IJKL}\left(\frac{x_{12}}{x_{13}},
\frac{x_{23}}{x_{13}} \right)
\ee
By Fourier transforming the above expression, we get the factorization 
\be
\langle \sigma^{I}_{\vk_1}\sigma^{J}_{\vk_2}\sigma^K_{\vk_{3}}\sigma^L_{\vk_{4}}\rangle'  \sim 
P_{\vk_4}\widetilde{F}^{IJKL}(k_{14},k_2,k_3)=
P_{\vk_4}\widetilde{F}^{IJKL}(k_{1},k_2,k_3), \label{sa}
\ee
where in the last equality we have used the fact that $k_4\ll k_1,k_2,k_3$. Then, the function $F^{IJKL}$ depends on just three momenta
and conformal invariance specifies this function to be necessarily proportional to the three point function.  
In particular, in order to calculate the proportionality factor in (\ref{sa}) one can make   use of the OPEs of the scalar fields. Let us see how this works in a particular model where the light fields mix through a quartic interaction of the type
\be
V_{\rm int}\sim \lambda_{IJKL} \sigma^I\sigma^J\sigma^K\sigma^L. \label{v4}
\ee
The OPE for the two fields $\sigma^I$  
and $\sigma^J$    in the (12) channel at the coincident point is 
\be
&&\sigma^I(\vx_1)\sigma^J(\vx_2)\!=\!\frac{C_0^{IJ}(w)}{x_{12}^{2w}}+
\frac{{C^{IJ}}_M(w)}{x_{12}^w}\sigma^M(\vx_2) +\sigma^I(\vx_2)\sigma^J(\vx_2)+\cdots.
\label{12}
\ee
Multiplying both sides with $\sigma^K(\vx_3)$ and since $x_{13}\approx 0$, we get
\be
&&\sigma^I(\vx_1)\sigma^J(\vx_2)\sigma^K(\vx_3)=\frac{C_0^{IJ}(w)}{x_{12}^{2w}}\sigma^K(\vx_3)+
\frac{{C^{IJ}}_M(w)}{x_{12}^w}\sigma^M(\vx_2)\sigma^K(\vx_3)\nonumber \\
&&\hspace{3.5cm}+\Big{(}\sigma^I(\vx_2)\sigma^J(\vx_2)\Big{)}\sigma^K(\vx_3)+\cdots.  \label{123}
\ee
For the operator ${\cal{O}}^{IJ}(\vx_2)=\sigma^I(\vx_2)\sigma^J(\vx_2)$ we have the OPE
\be
{\cal{O}}^{IJ}(\vx_2)\sigma^K(\vx_3)=\frac{D^{IJK}}{x_{23}^{3w}}+\frac{\lambda^{IJKL}}{x_{23}^{2w}}\sigma^L(\vx_3)+\cdots
\label{os}
\ee
as we assume that the operator ${\cal{O}}^{IJ}\sigma^K$ mixes with $\sigma^L$ through the quartic interaction. This is due to in field
theory is called mixing of composite operators.
By using Eqs.  (\ref{12}), (\ref{os}) in Eq. (\ref{123}) we then get

\be
&&\sigma^I(\vx_1)\sigma^J(\vx_2)\sigma^K(\vx_3) =
\frac{{C^{IJ}}_M}{x_{12}^w}\frac{C_0^{MK}}{x_{12}^wx_{23}^{w}}+\frac{D^{IJK}}{x_{23}^{3w}}+\frac{C_0^{IJ}(w)}{x_{12}^{2w}}\sigma^K(\vx_3)\\
&&\hspace{3.5cm} +\frac{{C^{IJ}}_M}{x_{12}^w}
\frac{{C^{MK}}_N(w)}{x_{23}^w}\sigma^N(\vx_3)+\frac{{\lambda^{IJK}}_N}{x_{23}^{w}x_{13}^w}\sigma^N(\vx_3)\cdots,
\label{1233}
\ee
where for symmetry reasons we have replaced $x_{23}^{2w}$ with $x_{13}^w x_{23}^w$.
Then, it is easy to find that after multiplying the above OPE with $\sigma^L(\vx_4)$, the connected part of the four-point 
correlator turns out to be
\be
\langle \sigma^I(\vx_1)\sigma^J(\vx_2)\sigma^K(\vx_3)\sigma^L(\vx_4)\rangle=\frac{D^{IJKL}}{x_{12}^wx_{23}^wx_{34}^{2w}} +
\rm{permutations},
\label{4pt}
\ee
where 
\be
\fbox{$\displaystyle
D^{IJKL}=\lambda^{IJKL}+{C^{IJ}}_MC^{MKL}$}. \label{4ptd}
\ee
By Fourier transforming the expression (\ref{4ptd}) we may get the four-point function in momentum space 
\be
&&\langle \sigma_{\vk_1}^I\sigma_{\vk_2}^J\sigma_{\vk_3}^K\sigma_{\vk_4}^L\rangle^\prime \sim D^{IJKL}C^{-1}_{NPM}
\,P_{\vk_4}\,
 \langle \sigma_{\vk_1}^N\sigma_{\vk_2}^P\sigma_{\vk_{34}}^M\rangle^\prime\nonumber \\
&&\hspace{2.7cm}
\sim D^{IJKL}C^{-1}_{NPM}
\,P_{\vk_4}\,
 \langle \sigma_{\vk_1}^N\sigma_{\vk_2}^P\sigma_{\vk_3}^M\rangle^\prime
,\,\,\,\,\,\,\,\,\,\,\,\,\,\,\,\,\,\,\,
(k_4\ll k_1,k_2,k_3).
\label{4pc0}
 \ee
Therefore, we confirm again that the  non-universal contribution to the four-point correlator in the squeezed  
limit does not have the  shape of the universal contribution. 

\subsection{Second method:  Cardy's trick}
For completeness, let us reproduce 
the expression (\ref{4oplongt})  
by making use of   the conformal 
symmetries of de Sitter. 
In order to apply the OPE expansion, we need the operators to be at almost coincident points. In our case,
three points are very close  and one point is far form the others. So, to make use of the OPE's,  we should bring the remote point $\vx_4$
very close to the points $\vx_1$,$\vx_2$ and $\vx_3$. To do this, we use the following trick due to Cardy \cite{cardy}.   We use the  fact that under the conformal inversion around an arbitrary point $\vx_0$, the conformal inversion (\ref{inv}) becomes

\be
\vx'=\frac{\vx-\vx_0}{|\vx-\vx_0|^2}.
\ee
In other words, if the point $\vx_4$ is very far from the point $\vx_0$, then under conformal inversion it becomes very close to it.
 Under the conformal inversion the four-point correlator transforms as 
\be
\langle \sigma^I(\vx_1)\sigma^J(\vx_2)\sigma^K(\vx_3)\sigma^L(\vx_4)\rangle=
\frac{\langle \sigma^I(\vx_1')\sigma^J(\vx_2')\sigma^K(\vx_3')\sigma^L(\vx_4')\rangle}{ |\vx_1-\vx_0|^{2w}|\vx_2-\vx_0|^{2w}|\vx_3-\vx_0|^{2w}|\vx_4-\vx_0|^{2w}}. \label{prod}
\ee
In our case, we may do a conformal inversion about the point $\vx_0\equiv\vx_3$. Then, the point $\vx_3$ will remain invariant
$(\vx_3'=\vx_3)$ and the point $\vx_4$ will come close to $\vx_3$ as $
x'_{34}=1/x_{34}\rightarrow 0$. All other distances will be clearly larger that 
$x'_{34}$. Thus, we may use the OPE 
\be
\sigma^K(\vx_3)\sigma^L(\vx'_4)=
\frac{\delta^{KL}}{x_{34}^{'2w}}+
\frac{C^{KLM}(g)}{x_{34}^{'w}}\sigma^M(\vx_3)+\sigma^K(\vx_3)\sigma^L(\vx_3)\cdots, \label{3sa}
\ee
where by $g$ we again denote collectively the couplings. We get therefore  from the linear term in the above OPE
 the connected part
\be
\langle \sigma^I(\vx_1')\sigma^J(\vx_2')\sigma^K(\vx_3)\sigma^L(\vx_4')\rangle=
\frac{C^{KLM}}{x_{34}^{'w}}\frac{\langle \sigma^I(\vx_1')\sigma^J(\vx_2')
\sigma^M(\vx_3)\rangle}{ |\vx_1-\vx_0|^{2w}|\vx_2-\vx_0|^{2w}|\vx_3-\vx_0|^{2w}|\vx_4-\vx_0|^{2w}}.
\ee
Now,  we  go back to the original configuration by doing again a second conformal inversion about the $\vx_3$ point. 
Since the conformal inversion will cancel three terms of the product in Eq. (\ref{prod}), we   
get  
\be
\langle \sigma^I(\vx_1)\sigma^J(\vx_2)\sigma^K(\vx_3)\sigma^L(\vx_4)\rangle
=\frac{C^{KLM}}{x_{34}^{w}} \langle \sigma^I(\vx_1)\sigma^J(\vx_2)\sigma^M(\vx_3)\rangle.
\ee
By Fourier transforming the above expression we finally obtain the second piece of Eq. (\ref{4oplongt})
\be
\langle \sigma^{I}_{\vk_1}\sigma^{J}_{\vk_2}\sigma^K_{\vk_{3}}\sigma^L_{\vk_{4}}\rangle'\supset {C^{KL}}_M P_{\vk_4}
\langle\sigma^{I}_{\vk_1}\sigma^{J}_{\vk_2}\sigma^M_{\vk_{4}}\rangle',\,\,\,\,\,\,(k_4\ll k_1,k_2,k_3).
\label{4opt}
\ee
We should stress again that $C^{KLM}$ should be understood in a non-perturbative sense and have a perturbative 
expansion in the coupling constants $g$, of the form  $C^{IJK}(g)=\sum_n c^{IJK}_n g^n$. 

 In fact, it is easy to generalize the above discussion to 
the general $n$-point function in the squeezed limit. A simple induction of the above 
leads to 
\be
\langle \sigma^{I}_{\vk_1}\sigma^{J}_{\vk_2}\cdots\sigma^K_{\vk_{n-1}} \sigma^L_{\vk_{n}}\rangle' \supset C^{KLM} P_{\vk_n}
\langle\sigma^{I}_{\vk_1}\sigma^{J}_{\vk_2}\cdots\sigma^M_{\vk_{n-1}}\rangle' ,\,\,\,\,\,\,(k_n\ll k_1,k_2,\cdots,k_{n-1}).
\ee
The first piece of  Eq. (\ref{4oplongt}) emerges from the quadratic term of the OPE in Eq. (\ref{3sa}). 
Indeed, 
this  term gives a contribution of the form
\be
\langle \sigma^I(\vx_1')\sigma^J(\vx_2')\sigma^K(\vx_3)\sigma^L(\vx_4')\rangle\supset 
\frac{\langle \sigma^I(\vx_1')\sigma^J(\vx_2')\sigma^K(\vx_3)\sigma^L(\vx_3)\rangle}{
 |\vx_1-\vx_0|^{2w}|\vx_2-\vx_0|^{2w}|\vx_3-\vx_0|^{2w}|\vx_4-\vx_0|^{2w}}.
\ee
For the operator ${\cal{O}}^{KL}(\vx_3)=\sigma^K(\vx_3)\sigma^L(\vx_3)$ we have the OPE
\be
{\cal{O}}^{KL}(\vx_3)\sigma^J(\vx_2')=\frac{D^{IJK}}{|\vx_2'-\vx_3|^{3w}}+
\frac{\lambda^{IJKL}}{|\vx_2'-\vx_3|^{2w}}\sigma^L(\vx_3)+\cdots
\label{os}
\ee
where we assume again that  ${\cal{O}}^{KL}\sigma^J$ mixes with $\sigma^L$ through the quartic interaction (\ref{v4}).
Repeating the steps above and after an inverse conformal inversion we get 
%
\be
\langle \sigma^I(\vx_1)\sigma^J(\vx_2)\sigma^K(\vx_3)\sigma^L(\vx_4)\rangle\supset
\frac{\lambda^{IJKL}}{x_{12}^wx_{23}^wx_{34}^{2w}} +
\rm{permutations},
\label{4pt}
\ee
Therefore the total four-point correlator (\ref{4oplongt}) may be written after Fourier transforming as in Eq. (\ref{4pc0})
where $D^{IJKL}$ is given by Eq. (\ref{4ptd}).  

\section{On the non-universal contributions to NG}
Having established the form of the non-universal three- and four-point correlator in the squeezed limit, we now focus our attention on their magnitude. We already pointed out that the non-universal contributions can be the dominant ones. For practical purposes, 
let us consider  the squeezed limit of the three-point correlator in a $V(\sigma)=\lambda \sigma^4/4!$ model

\be
\langle\sigma_{\vk_1}\sigma_{\vk_2} \sigma_{\vk_3}\rangle'\simeq-\frac{2}{3} \frac{\lambda\overline{\sigma}}{H^{2}} \, N_{k_t}\, P_{\vk_1}P_{\vk_2},\,\,\,\,\,\,\,\,\,\,\,\,\,\,\,\,\,\,\,\,\,\,\,\,\,\,\,\,\,(k_1\ll k_2\sim k_3).
\ee  
The corresponding contribution to the three-point correlator of the comoving curvature perturbation is 

\be
\langle\zeta_{\vk_1}\zeta_{\vk_2} \zeta_{\vk_3}\rangle'\simeq-\frac{ 2}{3N^{\prime }} \frac{\lambda\overline{\sigma}}{H^{2}}  \, N_{k_t}\, P^\zeta_{\vk_1}P^\zeta_{\vk_2},\,\,\,\,\,\,\,\,\,\,\,\,\,\,\,\,\,\,\,\,\,\,\,\,\,\,\,\,\,(k_1\ll k_2\sim k_3)
\ee  
leading to a non-universal contribution to the nonlinear parameter $f_{\rm NL}$ given by

\be
f_{\rm NL}^{\rm n-un}\simeq -\frac{5}{18 N^{\prime }} \frac{\lambda\overline{\sigma}}{H^{2}} N_{k_t}=-\frac{5}{36\pi} \frac{\lambda\overline{\sigma}}{H} \frac{N_{k_t}}{{\cal P}_\zeta^{1/2}}\simeq -\frac{5 \sqrt{2\lambda}}{36\pi} \frac{m(\overline{\sigma})}{H} \frac{N_{k_t}}{{\cal P}_\zeta^{1/2}}\simeq -65\left(\frac{\lambda}{10^{-2}}\right)^{1/2}\left(\frac{m(\overline{\sigma})/H}{10^{-2}}\right) \left(\frac{N_{k_t}}{50}\right),\nonumber\\
&&
\end{eqnarray}
where we have defined the quantity ${\cal P}_\zeta=(k^3/2 \pi^2)P^\zeta_{\vk}=N^{\prime 2}(H/2\pi)^2\simeq 2.3\cdot 10^{-9}$ and the (field dependent) Higgs mass $m(\overline{\sigma})^2=\lambda\overline{\sigma}^2/2$. Analogously we find

\begin{eqnarray}
g_{\rm NL}^{\rm n-un}\simeq -\frac{25}{54\cdot 3 N^{\prime 2 }} \frac{\lambda}{H^{2}} N_{k_t}=-\frac{25\lambda}{54\cdot 3 (2\pi)^2}  \frac{N_{k_t}}{{\cal P}_\zeta}\simeq - 10^6\left(\frac{\lambda}{10^{-2}}\right) \left(\frac{N_{k_t}}{50}\right).
\end{eqnarray}
We see that the nonlinearities generated by the non-universal pieces can be substantial, confirming previous findings \cite{zal,bmr,fr1,fr2}.
Furthermore, we also deduce that it is quite simple to have models in which $g_{\rm NL}\gsim f_{\rm NL}^2$ (in absolute values), a simple $\lambda\sigma^4$ model will do it as long as $H/\overline{\sigma}\gsim \sqrt{\lambda}$.

\section{Conclusions}
\noindent
In this paper we have made use of the OPE technique and,  partly, of the symmetries of the de Sitter epoch, to characterize the NG four-point correlator of the curvature perturbation in multifield inflation. In particular we have pointed out that

\begin{itemize}

\item The contribution to the squeezed limit of the four-point correlator coming from the intrinsic NG of the light fields at horizon crossing (which we dubbed non-universal) can be larger
than the superhorizon contributions (we dubbed them universal) generated even if the light fields are gaussian.

\item The shape of the non-universal contribution to the squeezed limit of the four-point correlator can be expressed in terms of  the three-point correlator. Nevertheless, in general the shapes of the universal and non-universal contributions
are different as the three-point correlator is note evaluated in the squeezed limit.

\end{itemize}
Therefore, particular care needs to be taken when studying the effects of the primordial NG on real observables, {\it e.g.}  the scale dependence
of the local halo bias in the presence of NG, as the squeezed limit of the four-point correlator is not expressible in terms of products of power spectra. The consequences of our findings will be investigated elsewhere.

\section*{Acknowledgments}
We thank C. Byrnes M. Sloth for interesting comments on the draft.  A.R. is supported by the Swiss National
Science Foundation (SNSF), project `The non-Gaussian Universe" (project number: 200021140236).


\end{document}